\documentclass[twocolumn,runningheads]{svjour2}
\smartqed  
\usepackage{graphicx}
\journalname{Astrophysics and Space Science}
\begin{document}

\title{Results from MAGIC's first observation cycle on galactic sources}



\author{Javier Rico (for the MAGIC Collaboration)}

\authorrunning{J. Rico} 

\institute{Javier Rico \at
              Institut de F\'{\i}sica d'Altes Energies \\
              Edifici Cn Universitat Aut\`onoma de Barcelona\\
              08193 Bellaterra (Barcelona) Spain \\
              Tel.: +34 93 581 3551\\
              Fax: +34 93 581 1938\\
              \email{jrico@ifae.es}           
}

\date{Received: date / Accepted: date}

\maketitle

\begin{abstract}
During its Cycle I, the MAGIC telescope targeted about 250 hours
several galactic sources sought to be, or detected previously by other
experiments in the same energy domain, $\gamma$-ray emitters. This
paper reviews some results of such MAGIC observations covering, among
others, supernova remnants, the Galactic Center and microquasars. We
will concentrate on the recent discovery at very high energy
$\gamma$-rays of the microquasar LS~I~+61~303.
\end{abstract}

\section{Introduction: The MAGIC Telescope} 
\label{sec:intro}
MAGIC is a telescope for very high energy (VHE, $E \geq 50-100$~GeV)
$\gamma$-ray observation exploiting the Imaging Air Cherenkov (IAC)
technique. It is located on the Roque de los Muchachos Observatory
($28^\circ 45^\prime 30^{\prime\prime}$N, $17^\circ$ $52^\prime$
$48^{\prime\prime}$W, 2250~m above see level) in La Palma (Spain). This
kind of instrument images the Cherenkov light produced in the particle
cascade initiated by a $\gamma$-ray in the atmosphere. MAGIC
incorporates a number of technological improvements in its design and
is currently the largest single-dish telescope (diameter 17~m) in this
energy band, yielding the lowest threshold ($\sim$50~GeV). It is
equipped with a 576-pixel photomultiplier camera with a 3.5$^\circ$
field of view. MAGIC's sensitivity above 100 GeV is $\sim 2.5\%$ of
the Crab nebula flux (the calibration standard candle for IAC
telescopes) in 50 hours of observations. The energy resolution above
200~GeV is better than 30$\%$. The angular resolution is $\sim
0.1^\circ$, while source localization in the sky is provided with a
precision of $\sim 2'$. MAGIC is also unique among IAC telescopes by
its capability to operate under moderate illumination (i.e.\ moonlight
and twilight). This allows to increase the duty cycle by a factor 1.5
and a better sampling of variable sources is possible.

The physics program developed with the MAGIC telescope includes both,
topics of fundamental physics and astrophysics. In this paper we
present the results regarding the observations of galactic
targets. The results from extragalactic observations are presented
elsewhere in these proceedings~\cite{mazin}.

\section{Highlights of cycle I} 
\label{sec:highlights}

MAGIC's first observation cycle spanned the period from January 2005
to April 2006. About 1/4 of the observation time (not counting that
devoted to Crab nebula technical observations) was devoted to galactic
objects. The observations covered both, candidates and well
established VHE $\gamma$-ray emitters, and included the following
types of objects: supernova remnants (SNRs), pulsars, pulsar wind
nebulae (PWN), microquasars ($\mu$QSRs), the Galactic Center (GC), one
unidentified TeV source and one cataclysmic variable. In this section
we highlight the results obtained so far from such observations, and
concentrate, in section~\ref{sec:lsi}, on the most interesting case of
the microquasar LS~I~+61~303.

\subsection{The Crab nebula and pulsar}

The Crab nebula is a steady emitter at GeV and TeV energies, what
makes it into an excellent calibration candle. The Crab nebula has
been observed extensively in the past over a wide range of
wavelengths, covering the radio, optical and x-ray bands, as well as
high-energy regions up to nearly 100 TeV. Nevertheless, some of the
relevant physics phenomena are expected to happen in the VHE domain,
namely the spectrum showing an inverse Compton (IC) peak close to
100~GeV, a cut-off of the pulsed emission somewhere between 10 and 100
GeV, and the verification of the flux stability down to the percent
level. The existing VHE $\gamma$-ray experimental data is very well
described by electron acceleration followed by the IC scattering of
photons generated by synchrotron radiation (synchrotron self Compton
process). Probing the presence/absence of a small contribution of VHE
$\gamma$'s produced in hadronic interactions is a challenge for
experimenters.

\begin{figure}
\centering
\includegraphics[width=8.5cm]{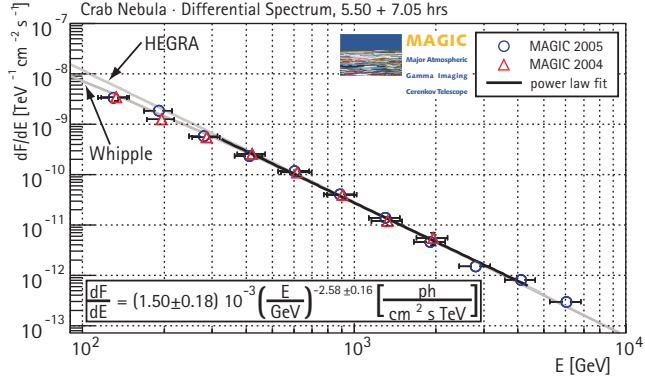}
\caption{Energy spectrum above 100~GeV from the Crab nebula measured
by MAGIC in two different observation seasons. For comparison, the
extrapolations down to 100 of HEGRA and Whipple measurements are shown.}
\label{fig:crab}
\end{figure}

Along the first cycle of MAGIC's regular observations, a significant
amount of time has been devoted to observe the Crab nebula, both for
technical and astrophysical studies. The performance of the telescope
has been experimentally evaluated and found in good agreement with the
expectations and Monte Carlo
simulations~\cite{wagner,performance}. This has allowed us to perform
routine analyses above 100~GeV, where the performance of our
instrument is fully understood. On the other hand, a sample of
12~hours of selected data has been used to measure with high precision
the spectrum down to $\sim$100~GeV, as shown in
figure~\ref{fig:crab}~\cite{wagner}. We have also carried out a search
for pulsed $\gamma$-ray emission from Crab pulsar and two millisecond
pulsars~\cite{lopez,ona}, albeit without positive result. The derived
upper limits of the pulsed flux for the three observed pulsars are
shown in Table~\ref{table:pulsar}.

\begin{table}
\caption{Upper limits (UL) to the pulsed integral
flux from three pulsars observed by MAGIC. The considered energy
threshold ($E_\mathrm{th}$), observation time (OT), assumed duty cycle
(DC) and confidence level (CL) are also shown.}
\centering
\small
\label{table:pulsar}   
\begin{tabular}{lccccc}
\hline\noalign{\smallskip}
Pulsar & $E_\mathrm{th}$  & OT & DC  & CL & UL  \\[3pt]
 & \tiny[GeV] & \tiny[hour] & \tiny($\%$) & \tiny($\%$) &\tiny[ph s$^{-1}$cm$^{-2}$] \\[3pt]
\tableheadseprule\noalign{\smallskip}
Crab           &  90 & 4  & 21 & 95 & 2.0$\times 10^{-10}$ \\
Crab           & 150 & 4  & 21 & 95 & 1.1$\times 10^{-10}$ \\
PSR B1957+20   & 115 & 6  & 5  & 90 & 2.9$\times 10^{-11}$ \\
PSR J0218+4232 & 115 & 24 & 5  & 90 & 1.1$\times 10^{-10}$ \\
\noalign{\smallskip}\hline
\end{tabular}
\normalsize
\end{table}

\subsection{Supernova Remnants}

Shocks produced at supernova explosions are assumed to be the source
of the galactic component of the cosmic ray flux~\cite{zwicky}. The
proof that this is the case could be provided by observations in the
VHE domain. The rationale is that the hadronic component of the cosmic
rays --enhanced close to their source, i.e.\ the SNR-- should produce
VHE $\gamma$-rays by the interaction with nearby dense molecular
clouds. Although recent data seem to indicate that this is the case,
it is difficult to disentangle the VHE component initiated by
hadrons from that produced by Bremsstrahlung and IC processes by
accelerated electrons. Therefore more data in the TeV regime together
with multi-wavelength studies are needed to finally solve the
long-standing puzzle of the origin of galactic cosmic rays.

\begin{figure}
\centering
\includegraphics[width=8.7cm]{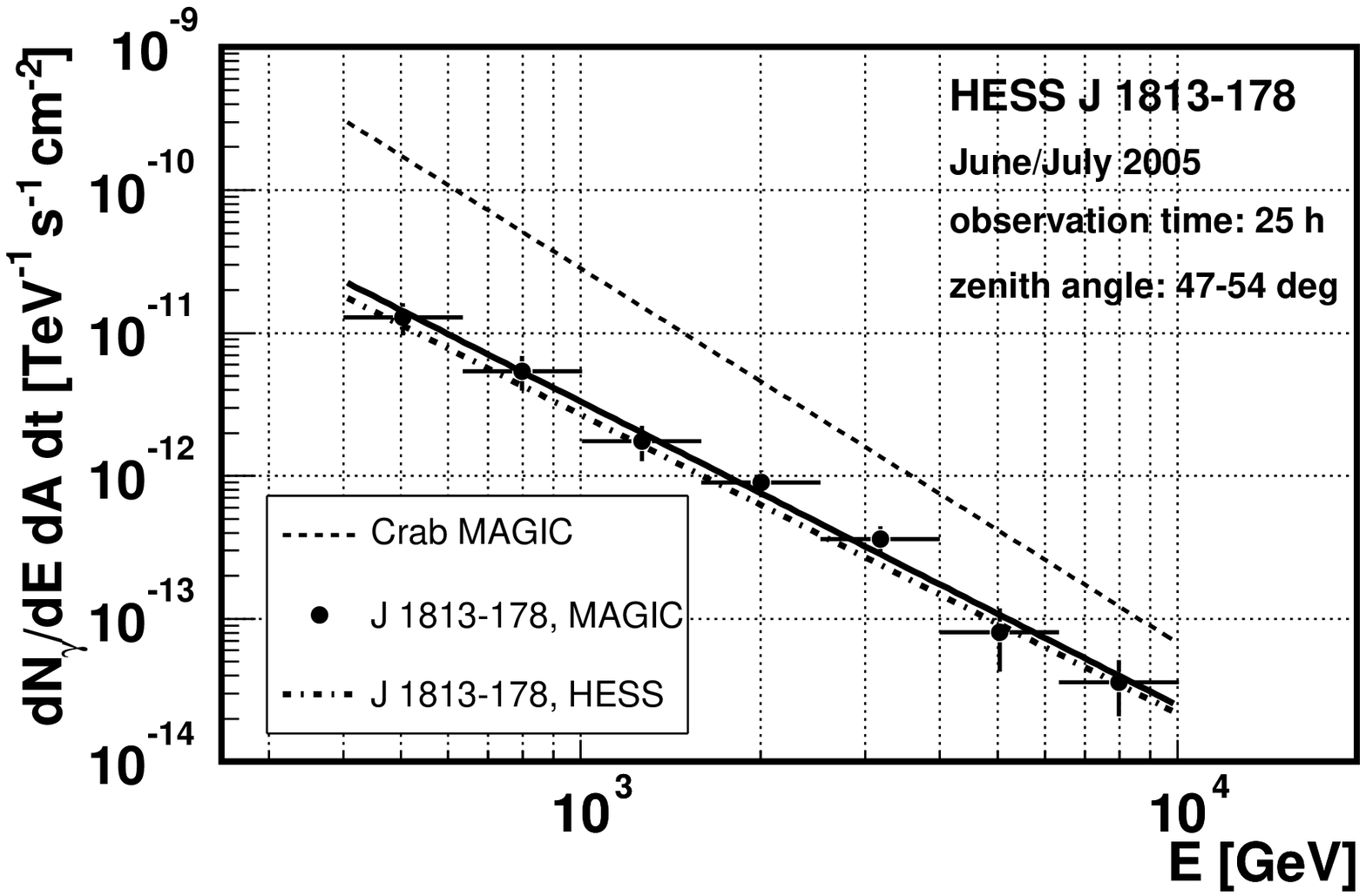}
\includegraphics[width=8.5cm]{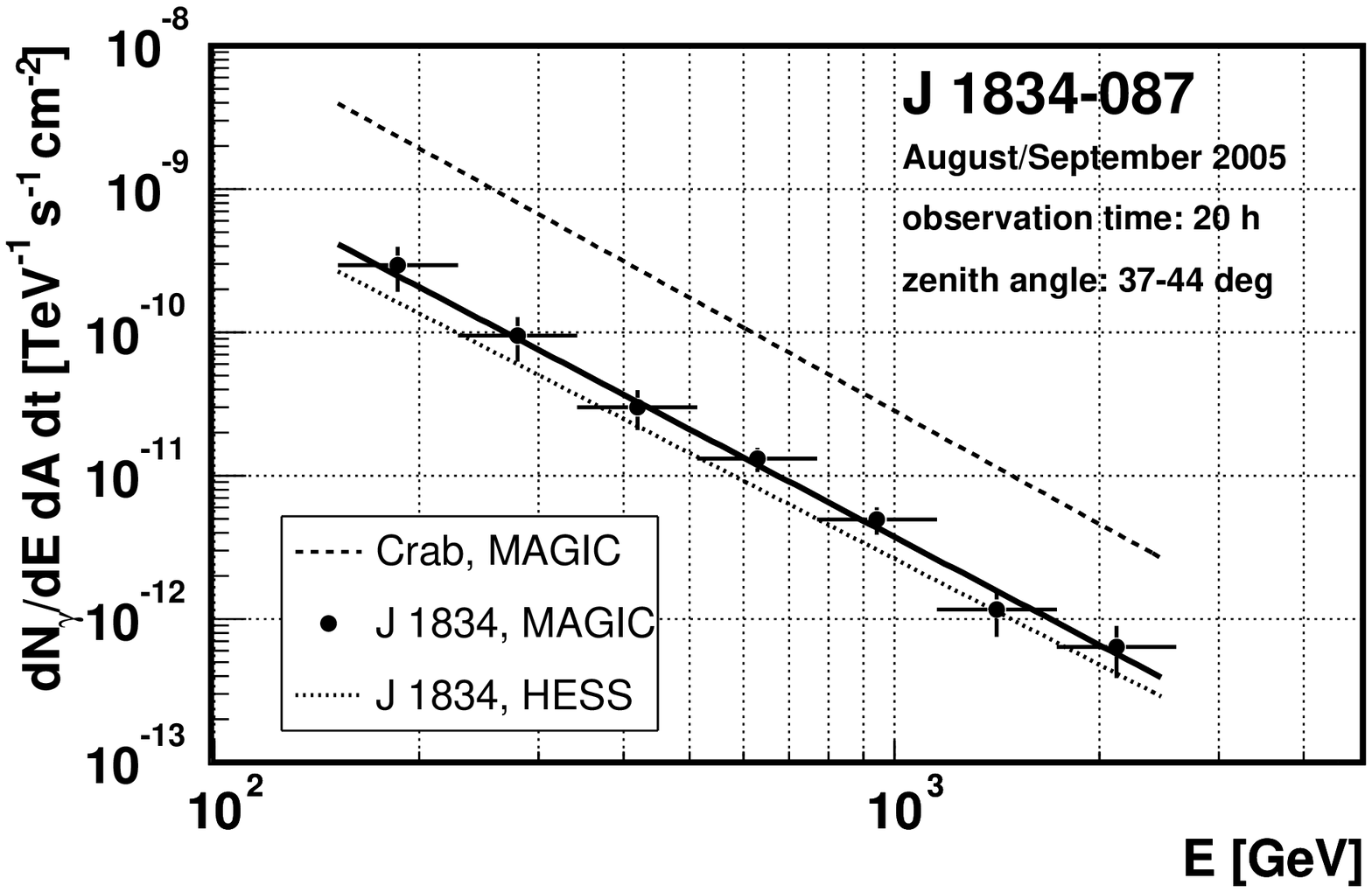}
\caption{Energy spectrum above 400~GeV of HESS~J1813-178 (top)
and above 150 GeV of HESS~J1834-087 (bottom), measured by MAGIC. For
comparison, also the spectra measured by HESS and that of the Crab
nebula are shown.}
\label{fig:snr}
\end{figure}

\begin{figure}
\vspace{0.3cm}
\centering
\includegraphics[width=8.5cm]{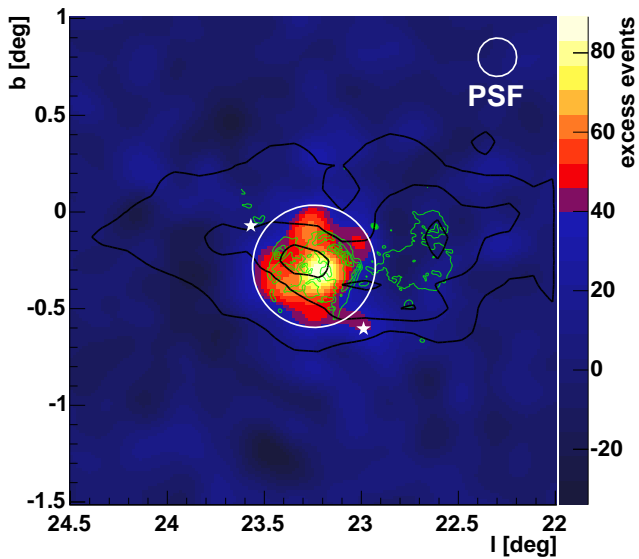}
\caption{Sky map of $\gamma$-ray candidate events (background
subtracted) in the direction of HESS~J1834-087 for an energy
threshold of about 250 GeV. Overlayed are $^{12}$CO emission contours
(black)~\cite{dame} and contours of 90~cm VLA radio data (green)~\cite{white}.}
\label{fig:hess1834}
\end{figure}

Within its program of observation of galactic sources, MAGIC has
observed a number of supernova remnants. In particular, we are
observing several of the brightest EGRET sources associated to SNRs,
and the analysis of the acquired data is in progress. On the other
hand, we have confirmed the VHE $\gamma$-ray emission from the SNRs
HESS~J1813-178~\cite{hess1813} and HESS~J1834-087
(W41)~\cite{hess1834}. Our results have confirmed SNRs as a well
established population of VHE $\gamma$-ray emitters. The energy
spectra measured by MAGIC for these two sources are shown in
Figure~\ref{fig:snr}. They are, both, well described by an unbroken
power law and an intensity of about 10$\%$ of the Crab nebula
flux. Furthermore, MAGIC has proven its capability to study moderately
extended sources by observing HESS~J1834-087. The morphology of this
object measured by MAGIC is shown in
Figure~\ref{fig:hess1834}. Interestingly, the maximum of the VHE
emission has been correlated with a maximum in the density of a nearby
molecular cloud (shown in the figure by the contour lines of the
$^{12}$CO emission intensities). Although the mechanism responsible
for the VHE radiation remains yet to be clarified, this is a hint that
it could be produced by high energy hadrons interacting with the
molecular cloud.

\subsection{Galactic Center}

\begin{figure}
\vspace{0.5cm}
\centering
\includegraphics[width=8.5cm]{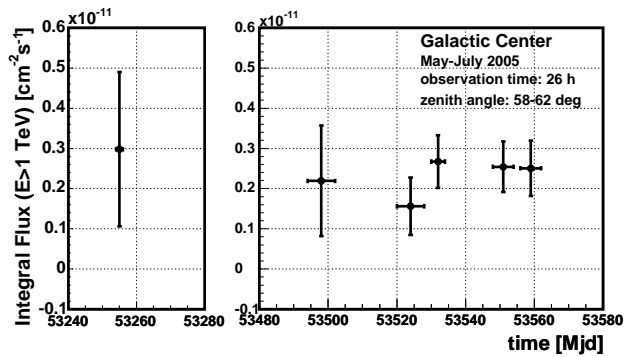}
\caption{Integral VHE $\gamma$-ray flux above
1 TeV as a function of time from the GC as measured by MAGIC during
September 2004 and May-July 2005.}
\label{fig:sgra}
\end{figure}

We have also measured the VHE $\gamma$-ray flux from the
GC~\cite{sgra}. The possibility to indirectly detect dark matter
through its annihilation into VHE $\gamma$-rays has risen the interest
to observe this region during the last years. Our observations have
confirmed a point-like $\gamma$-ray excess whose location is spatially
consistent with Sgr A* as well as Sgr A East. The energy spectrum of
the detected emission is well described by an unbroken power law of
photon index $\alpha=-2.2$ and intensity about $10\%$ of that of the
Crab nebula flux at 1~TeV. This result disfavours dark matter
annihilation as the main origin of the detected flux. Furthermore,
there is no evidence for variability of the flux on hour/day time
scales nor on a year scale, as shown in Figure~\ref{fig:sgra}. This
suggests that the acceleration takes place in a steady object such as
a SNR or a PWN, and not in the central black hole.

\subsection{Other observation targets}

MAGIC observation program of galactic sources includes also other
kinds of objects such as PWN, cataclysmic variables, unidentified
sources and $\mu$QSRs (both high and low mass). The analyses regarding
these sources are ongoing and will be reported in the near future. The
one exception is the high-mass x-ray binary LS~I~+61~303, discussed in
the section~\ref{sec:lsi}.

\section{The microquasar LS~I~+61~303} 
\label{sec:lsi}

\subsection{Microquasars}
Microquasars are a subclass of stellar, x-ray binary systems that
display prominent radio emission, usually attributed to the existence
of jets of relativistic particles.  They are named after the
similarities with active galactic nuclei (AGNs), since $\mu$QSRs show
the same three ingredients that make up radio-loud AGNs: a compact
object, an accretion disc, and relativistic
jets~\cite{mirabel}. Hence, $\mu$QSRs are galactic, scaled-down
versions of an AGN, where instead of a super-massive black hole we
deal with a compact object of just a few solar masses that accretes
material from a donor star.  The similarities with AGNs explain the
large interest risen by $\mu$QSRs. Crucial for our understanding of
accreting systems is also that: (i) $\mu$QSRs are nearby objects, and
(ii) they show very short timescale variability. Those reasons make
these objects ideal laboratories for the study of the physical
processes that govern how $\mu$QSRs and AGNs work. In particular, the
short timescale variability displayed by $\mu$QSRs allows to see
changes in the ongoing physical processes within typical time scales
ranging from minutes to months, in contrast with the usual scales of
years to observe such variability in AGNs. In addition, $\mu$QSRs
could measurably contribute to the density of galactic cosmic
rays~\cite{heinz}.

\subsection{LS~I~+61~303}
One of the most studied $\mu$QSR candidates is LS~I~+61~303. This
system is composed of a compact object of unknown nature (neutron star
or black hole) in a highly eccentric ($e=0.7$) orbit around a Be
star. The orbital period --with associated radio~\cite{gregoryold} and
x-ray~\cite{taylor} outbursts-- is 26.5 days and periastron passage is
at phase 0.23~\cite{casares}.  The phase and intensity of the radio
outburst show a modulation of 4.6
years~\cite{gregorynew}. High-resolution radio imaging techniques have
shown extended, radio-emitting structures with angular extension of
$\sim$0.01 to $\sim$0.1 arc-sec, interpreted within the framework of
$\mu$QSR scenario, where the radio emission is originated in a
two-sided, probably precessing, relativistic jet
($\beta/c=0.6$)~\cite{massi}. However, no solid evidence of the
presence of an accretion disk (i.e. a thermal x-ray component) has
been observed. LS~I~+61~303 is also one of the two $\mu$QSR candidates
positionally coincident with EGRET $\gamma$-ray
sources~\cite{kniffen}, and the only one located in the Northern
Hemisphere --hence a suitable target for MAGIC. There are also hints
of variability of the $\gamma$-ray flux~\cite{tavani}. However, the
large uncertainty of the position of the EGRET source has not allowed
an unambiguous association with LS~I~+61~303.

\subsection{MAGIC observations}
LS~I~+61~303 was observed in the VHE regime with MAGIC during 54 hours
(after standard quality selection, discarding bad weather data)
between October 2005 and March 2006~\cite{lsi}. The data analysis was
carried out using the standard MAGIC reconstruction and analysis
software~\cite{hess1813,hess1834,sgra}.

\begin{figure}[!t]
\centering
\includegraphics[width=7.5cm]{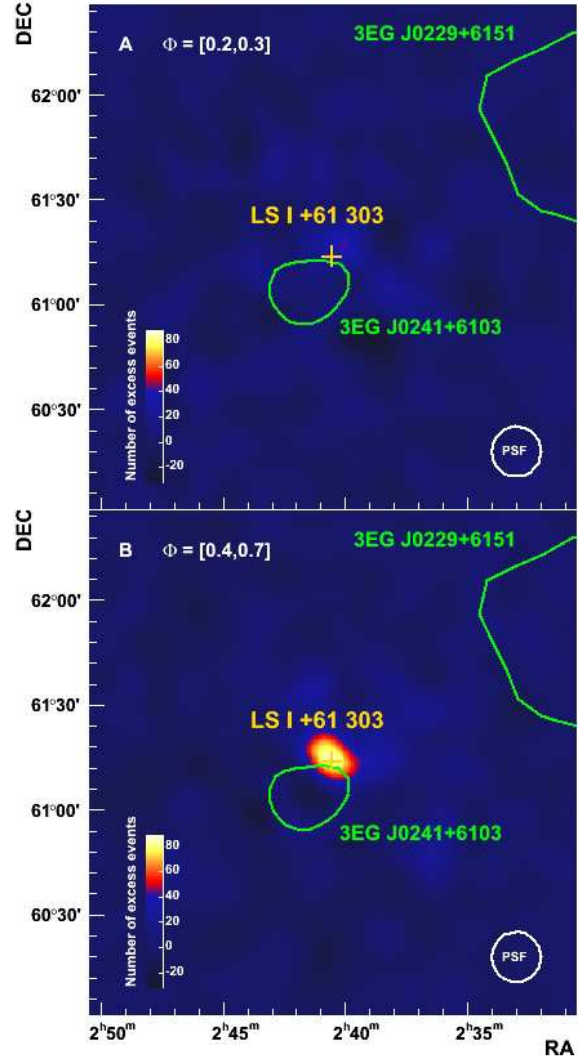}
\caption{
Smoothed maps of $\gamma$-ray excess events above 400 GeV around
LS~I~+61~303. (A) Observations over 15.5 hours corresponding to data
around periastron (i.e., between orbital phases 0.2 and 0.3). (B)
Observations over 10.7 hours at orbital phase between 0.4 and 0.7. The
number of events is normalized in both cases to 10.7 hours of
observation. The position of the optical source LSI~+61~303 (yellow
cross) and the 95$\%$ confidence level contours for the EGRET sources
3EG J0229+6151 and 3EG J0241+6103 (green contours) are also shown. The
bottom right circle shows the size of the point spread function of
MAGIC (1$\sigma$ radius). From Albert \emph{et al.}~\cite{lsi}.}
\label{fig:lsi-skymap}
\end{figure}

\begin{figure}
\vspace{0.5cm}
\centering
\includegraphics[width=8.5cm]{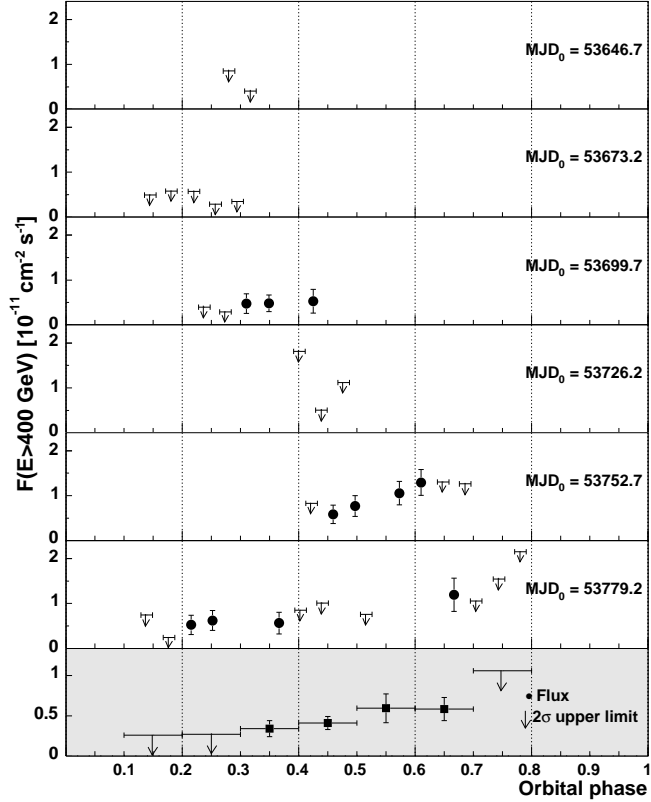}
\caption{VHE $\gamma$-ray flux of LS~I~+61~303 as a function of orbital
phase for the six observed orbital cycles (six upper panels, one point
per observation night) and averaged for the entire observation time
(bottom panel). Vertical error bars include 1$\sigma$ statistical error
and 10$\%$ systematic uncertainty on day-to-day relative fluxes. Only
data points with more than 2$\sigma$ significance are shown, and
2$\sigma$ upper limits~\cite{rolke} are derived for the rest. The modified
Julian date (MJD) corresponding to orbital phase 0 is indicated for
every orbital cycle. From Albert \emph{et al.}~\cite{lsi}.} 
\label{fig:lsi-lc}
\end{figure}

Figure~\ref{fig:lsi-skymap} shows the reconstructed $\gamma$-ray map
during two different observation periods, around periastron passage
and at higher (0.4-0.7) orbital phases. No significant excess in the
number of $\gamma$-ray events is detected around periastron passage,
whereas it shows up clearly (9.4$\sigma$ statistical significance) at
later orbital phases.  The distribution of $\gamma$-ray excess is
consistent with a point-like source and is located at (J2000):
$\alpha = 2^\mathrm{h}40^\mathrm{m}34^\mathrm{s}$, $\delta = 61^\circ
15^\prime 25^{\prime\prime}$, with statistical and systematic
uncertainties of $\pm 0.4^\prime$ and $\pm2^\prime$, respectively, in
agreement with the position of LS~I~+61~303. In the natural case in
which the VHE emission is produced by the same object detected at
EGRET energies, this result identifies a $\gamma$-ray source that
resisted classification during the last three decades.

\begin{figure}
\centering
\includegraphics[width=8cm]{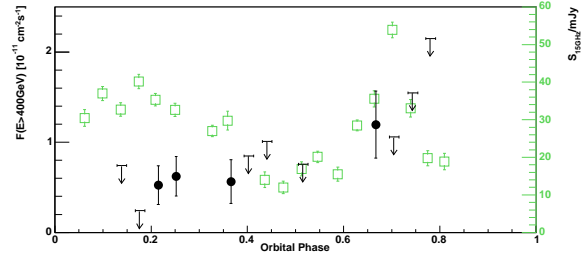}
\caption{LS~I~+61~303 radio flux density at 15~GHz measured with the
Ryle Telescope (green squares, right axis) and results from the last
orbital cycle observed by MAGIC (black dots, left axis), from 14
February to 8 March 2006. The MJD corresponding to orbital phase 0 is
indicated. From Albert \emph{et al.}~\cite{lsi}.}
\label{fig:lsi-ryle}
\end{figure}

Our measurements show that the VHE $\gamma$-ray emission from
LS~I~+61~303 is variable. The $\gamma$-ray flux above 400 GeV coming
from the direction of LS~I~+61~303 (see Figure~\ref{fig:lsi-lc}) has a
maximum corresponding to about 16$\%$ of the Crab nebula flux, and is
detected at around phase 0.6. The combined statistical significance of
the 3 highest flux measurements is 8.7$\sigma$, for an integrated
observation time of 4.2 hours. The probability for the distribution of
measured fluxes to be a statistical fluctuation of a constant flux
(obtained from a $\chi^2$ fit of a constant function to the entire
data sample) is $3\times 10^{-5}$. The fact that the detections occur
at similar orbital phases hints at a periodic nature of the VHE
$\gamma$-ray emission. Contemporaneous radio observations of
LS~I~+61~303 were carried out at 15 GHz with the Ryle Telescope
covering several orbital periods of the source. The peak of the radio
outbursts was at phase 0.7, i.e. between 1 and 3 days after the
increase observed at VHE $\gamma$-rays flux (see
Figure~\ref{fig:lsi-ryle}).

\begin{figure}
\vspace{0.5cm}
\centering
\includegraphics[width=8.5cm]{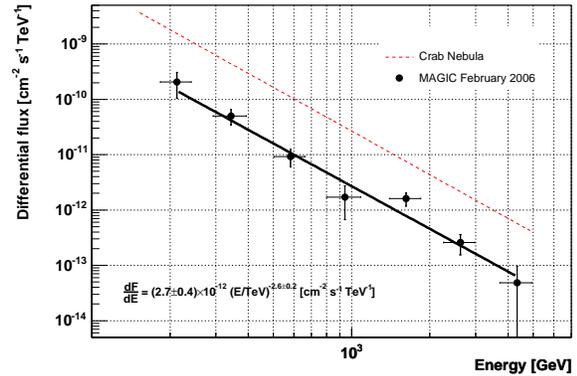}
\caption{
Differential energy spectrum of LS~I~+61~303 for energies
between 200 GeV and 4 TeV and averaged for orbital phases between 0.4
and 0.7, measured by MAGIC. The error bars show the 1$\sigma$
statistical uncertainty. The dashed, red line corresponds to the Crab
nebula differential spectrum also measured by MAGIC. The solid, black
line is a fit of a power law (also expressed mathematically in the
inset) to the measured points. From Albert \emph{et al.}~\cite{lsi}.}
\label{fig:lsi-spec}
\end{figure}

The VHE spectrum derived from data between $\sim$200 GeV and
$\sim$4~TeV at orbital phases between 0.4 and 0.7 (see
Figure~\ref{fig:lsi-spec}) is fitted reasonably well ($\chi^2/ndf =
6.6/5$) by a power law function:
\begin{eqnarray}
dN/(dA/dt/dE) & = & (2.7 \pm 0.4 \pm 0.8) \times 10^{-12} \times
\nonumber \\
& & E^{(-2.6 \pm 0.2 \pm
0.2)} \quad  \mathrm{cm}^{-2} \mathrm{s}^{-1} \mathrm{TeV}^{-1} 
\end{eqnarray}
where $N$ is the number of $\gamma$-rays reaching Earth per unit area
$A$, time $t$ and energy $E$ (expressed in TeV). Errors quoted are
statistical and systematic, respectively. This spectrum is consistent
with that measured by EGRET for a spectral break between 10 and
100~GeV. The flux from LS~I~+61~303 above 200~GeV corresponds to an
isotropic luminosity of $\sim 7 \times 10^{33}$~erg~s$^{-1}$ (assuming
a distance to the system of 2~kpc~\cite{frail}), of the same order of
that of the similar object LS~5039~\cite{hessls}, and a factor $\sim
2$ lower than the previous experimental upper limit ($< 8.8 \times
10^{-12}$~cm$^{-2}$ s$^{-1}$ above
500~GeV)~\cite{wipple}. LS~I~+61~303 displays more luminosity at GeV
than at x-ray energies, a behavior shared also by LS~5039.

\subsection{Emission scenarios}
LS~I~+61~303 belongs, together with LS~5039~\cite{hessls} and
PSR~B1259-63~\cite{1259}, to a new class of objects, the so-called
$\gamma$-ray binary systems, whose electromagnetic emission extends up
to the TeV domain. LS~I~+61~303 and LS~5039 are usually thought to be
$\mu$QSRs, since evidences of relativistic jets have been found at
radio frequencies. In such scenario, the high energy emission is
produced by the shocks triggered at the relativistic jets
\cite{mirabel}. However, no clear signal of the presence of an
accretion disk (in particular an spectral feature between $\sim 10$
and $\sim 100$ keV due to the cut-off of the thermal emission) has
been observed so far. Because of that, it has been alternatively
proposed that relativistic particles could be injected into the
surrounding medium by the wind from a young pulsar~\cite{maraschi},
which seems to be the case of PSR~B1259-63. In the case of
LS~I~+61~303 also the resemblance of the time variability and the
radio/x-ray spectra with those of young pulsars seem to support such
hypothesis. However, no pulsed emission has been detected from
LS~I~+61~303. Therefore, the existing data in the radio/x/$\gamma$-ray
domain, cannot conclusively confirm, or rule out, none of the two
proposed scenarios. Thus, the question of whether the three known
$\gamma$-ray binaries produce TeV emission by the same mechanism, and
by which one, is currently object of an intense
debate~\cite{perspectives}. Observation at VHE with MAGIC
simultaneously together with other instruments at other wavelength
domains --in particular radio-- will help elucidate the mechanism of
the TeV emission and hence the nature of $\gamma$-ray binaries.

The observation of jets has triggered the study of different
microquasar-based $\gamma$-ray emission models, some regarding
hadronic mechanisms: relativistic protons in the jet interact with
non-relativistic stellar wind ions, producing $\gamma$-rays via
neutral pion decay~\cite{romero}; some regarding leptonic mechanisms:
IC scattering of relativistic electrons in the jet on stellar and/or
synchrotron photons~\cite{valenti}.

The TeV flux measured by MAGIC has a maximum at phases 0.5-0.6 (see
Figure~\ref{fig:lsi-lc}), overlapping with the x-ray outburst and the
onset of the radio outburst. The timescale of the TeV variability
constrains the emitting region to be smaller than a few 10$^{15}$~cm
(or ~0.1 arc-sec at 2~kpc), which is larger than the binary system and
of the same order of the detected radio jet-like structures. The
maximum flux is not detected at periastron, when the accretion rate is
expected to be the largest. This result seems to favor the leptonic
over the hadronic models, since the IC efficiency is likely to be
higher than that of proton-proton collisions at the relatively large
distances from the companion star at this orbital phase. Further, VHE
$\gamma$-ray emission that peaks after periastron passage has been
predicted considering electromagnetic cascading within the binary
system~\cite{wlodek}. It is also possible to explain the orbital
modulation of the VHE emission in LS~I~+61~303 by the geometrical
effect of changes in the relative orientation of the stellar companion
with respect to the compact object and jet as it impacts the position
and depth of the $\gamma \gamma$ absorption trough~\cite{gupta}. The
existing data are, however, not conclusive to confirm or rule out any
of the theoretical models existing in the literature. Concerning
energetics, a relativistic power of several 10$^{35}$~erg~s$^{-1}$,
extracted from accretion, could explain the non thermal luminosity of
the source from radio to VHE $\gamma$-rays.

LS~I~+61~303 is an excellent laboratory to study the VHE $\gamma$-ray
emission and absorption processes taking place in massive x-ray
binaries: the high eccentricity of the binary system provides very
different physical conditions to be tested on timescales of less than
one month. Future MAGIC observations will test both, the periodicity
of the signal and its intra-night variability.

\paragraph{Acknowledgements.}
We thank the Instituto de Astrof\'{\i}sica de Canarias for the
excellent working conditions at the Observatory Roque de los Muchachos
in La Palma

\end{document}